\documentclass[journal=jacsat,manuscript=article]{achemso}

\usepackage{natbib,hyperref}       
\usepackage{url}                   
\usepackage{booktabs}              
\usepackage{amsfonts}              
\usepackage{amsmath}               
\usepackage{float}                 

\usepackage{xcolor}

\usepackage[version=3]{mhchem} 

\newcommand*{\angstrom}{\mbox{\normalfont\AA}}

\author{Filippo Balzaretti}
\email{filippo@hmi.uni-bremen.de}
\affiliation[HMI 1]
{Hybrid Materials Interfaces Group, Faculty of Production Engineering, University of Bremen, Bremen, Germany}
\alsoaffiliation[BCCMS 2]
{Bremen Center for Computational Materials Science, University of Bremen, Bremen, Germany}

\author{Verena Gupta}
\affiliation[CMS 3]
{Computational Materials Science Group, Faculty of Physics and Electrical Engineering, University of Bremen, Bremen, Germany}
\alsoaffiliation[BCCMS 2]
{Bremen Center for Computational Materials Science, University of Bremen, Bremen, Germany}

\author{Lucio Colombi Ciacchi}
\affiliation[HMI 1]
{Hybrid Materials Interfaces Group, Faculty of Production Engineering, University of Bremen, Bremen, Germany}
\alsoaffiliation[BCCMS 2]
{Bremen Center for Computational Materials Science, University of Bremen, Bremen, Germany}
\alsoaffiliation[MAPEX 4]
{MAPEX Center for Materials and Processes, University of Bremen, Bremen, Germany}

\author{Bálint Aradi}
\affiliation[CMS 3]
{Computational Materials Science Group, Faculty of Physics and Electrical Engineering, University of Bremen, Bremen, Germany}
\alsoaffiliation[BCCMS 2]
{Bremen Center for Computational Materials Science, University of Bremen, Bremen, Germany}

\author{Thomas Frauenheim}
\affiliation[CMS 3]
{Computational Materials Science Group, Faculty of Physics and Electrical Engineering, University of Bremen, Bremen, Germany}
\alsoaffiliation[BCCMS 2]
{Bremen Center for Computational Materials Science, University of Bremen, Bremen, Germany}
\alsoaffiliation[MAPEX 4]
{MAPEX Center for Materials and Processes, University of Bremen, Bremen, Germany}

\author{Susan K{\"o}ppen}
\email{koeppen@hmi.uni-bremen.de}
\affiliation[HMI 1]
{Hybrid Materials Interfaces Group, Faculty of Production Engineering, University of Bremen, Bremen, Germany}
\alsoaffiliation[BCCMS 2]
{Bremen Center for Computational Materials Science, University of Bremen, Bremen, Germany}
\alsoaffiliation[MAPEX 4]
{MAPEX Center for Materials and Processes, University of Bremen, Bremen, Germany}

\title
  {Water reactions on reconstructed rutile TiO$_2$: a DFT / DFTB approach}

\abbreviations{DFT, DFTB, AFM, STM, FPMD, SCF, DOS, SCC }

\DeclareUnicodeCharacter{2212}{-}
\DeclareUnicodeCharacter{2009}{\,} 
\DeclareUnicodeCharacter{0305}{ ̅ } 

\begin{document}
\begin{tocentry}

\begin{figure}[H]
\centering
\includegraphics{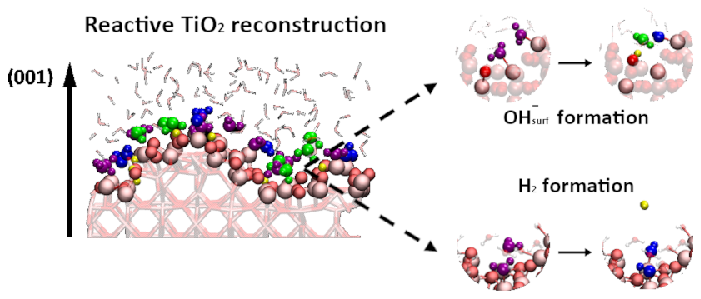}
\caption{reactive titania reconstructions lead to the formation of charged surface groups and molecular H$_2$}
\end{figure}

\end{tocentry}

\begin{abstract}

\noindent
Far from being conclusively understood, the reactive interaction of water with rutile does still present a challenge to atomistic modelling techniques rooted on quantum mechanics. We show that static geometries of stoichiometric TiO$_2$/water interfaces can be described well by Density Functional Tight Binding (DFTB). However, this method needs further improvements to reproduce the low dissociation propensity of H$_2$O after adsorption predicted by Density Functional Theory (DFT). A reliable description of the surface reactivity of water is fundamental to investigate the non-stoichiometric reconstruction of the (001) facet rich in Ti interstitials. Calculations based on (DFT) predict the transition temperature for the onset of reconstruction in remarkable agreement with experiments and suggest that this surface, in contact with liquid water, can promote spontaneous H$_2$O splitting and formation of H$_2$ molecules.

\end{abstract}

\section{Introduction}

The impact of titanium dioxide as a fundamental material for new technologies is worldwide recognized.
TiO$_2$ with negligible amount of impurities can be synthesized in the form of powder at very affordable prices~\cite{Alaa}, which has enabled its use in mass markets such as cosmetics or construction dyes. 
Moreover, its two main phases, anatase and rutile, are capable of major photocatalytic activity under UV light irradiation, and are thus employed in process-engineering applications such as water splitting~\cite{Migani, Beinik} or environmental cleaning~\cite{Lan, Dai, Ilina}. 
Anatase presents a larger electronic band gap than rutile (3.20~eV vs. 3.02~eV respectively~\cite{Monai}), which is advantageous for photo-catalysis because of the lower likelihood of electron-holes recombination~\cite{Odling}.
However, rutile is more stable thermodynamically, and thus offers better resistance to high temperatures and pressures~\cite{Hanaor, Zhang, Xiaobo, Kiarii, Fu}.
For this reason, the broad surface-science community has devoted much attention to the molecular features of both anatase and rutile surfaces, with the aim of elucidating the key structure-function relationships at the basis of their (photo)chemical properties~\cite{Diebold, Pang, Bourikas}. 
Experimental information about both pristine and defective surface features of anatase and rutile has been acquired mainly via atomic force microscopy (AFM) and scanning tunneling microscopy (STM)~\cite{Diebold}. 
Theoretical studies of the surface termination in vacuum, in oxygen-rich atmospheres or in the presence of water are based especially on Density Functional Theory (DFT)~\cite{Perron, Kiejna, Labat, Song}.

Despite the huge progress made in the past few decades, accurate atomistic models of TiO$_2$/water interfaces are still rare, with several contributions coming from our own work on crystalline oxides~\cite{Koeppen, Friedrichs1, Friedrichs2} and amorphous Ti oxidation layers~\cite{Schneider_2010, Schneider_2011}.
Especially the issue of surface reconstructions, which occur for several TiO$_2$ facets, still needs to be investigated in greater detail.
However, since such reconstructions result in much larger surface primitive cells, DFT-based predictions of the dynamical behaviour of interfacial water molecules quickly become extremely expensive from a computational point of view.

In the present work, we investigate the atomistic structure and relative thermodynamic stability of reconstructed rutile (100) and (001) surfaces, in comparison with the unreconstructed (110), (100) and (001) surfaces.
In particular, the reactivity of water on the reconstructed surfaces will be studied in both static and dynamic simulations.
While most electronic structure calculations and First-Principles Molecular Dynamics (FPMD) simulations will be performed at the full DFT level, we will also test the accuracy of a slightly modified set of parameters for self-consistent-charge Density Functional Tight Binding (DFTB).
Especially for larger-scale TiO$_2$/water interface models, we believe that a combination of DFT and DFTB might represent a viable path to tackle the (photo)chemical behaviour of realistic (nanosized) titanium oxide surfaces.
This is far from being a trivial task, as the two levels of theory shall reproduce in a satisfactory way both the structural geometries and the electronic energies in different situations.

Rutile (110) is the most stable and therefore most studied surface.
It has been a long disputed issue, up to the present day, to which extent water molecules adsorb dissociatively or molecularly on this surface.
Although some researchers assume that fully dissociative adsorption leading to terminal hydroxyl groups is the most favourable scenario  \cite{Zhao}, the more widely hold opinion is that mixed molecular/dissociative or entirely molecular adsorption takes place~\cite{Kumar, Liu, Agosta}.
In fact, theoretical investigations considering disordered oxide layers~\cite{Julian} or kink surface sites~\cite{Zheng} have suggested water dissociation to be promoted only by strongly undercoordinated Ti sites, bound to four or less oxygen atoms, and not by five-fold coordinated Ti atoms such as those present on rutile-(110). 
Different studies attributed this discrepancy to the fact that the adsorption mechanism depends on two main factors: the thickness of the simulated surface slab model and the coverage of water at the surface, with higher coverages or bulk liquid water stabilizing molecular adsorption (fully or at least partly)~\cite{Gonzales, Kamisaka, Harris}.  

In the common natural rutile powder, the (100) facet is present with a proportion of at least $20\%$ \cite{Perron}. 
At first glance, this surface looks quite different from rutile (110), although the coordination numbers of both O and Ti surface atoms are exactly the same in the two cases.
Upon annealing at more than $800$~K under UHV conditions, rutile (100) tends to reconstruct along (110) microfacets presenting a $1 \times 3$,  $1 \times 5$ or $1 \times 7$ surface unit cell with a roof-like shape (see Fig.~1)~\cite{Diebold, Zschack}.
Therefore, both the smooth and reconstructed surfaces have been studied regarding their atomic arrangement~\cite{Lindan, Sawai} and their interaction with adsorbate molecules~\cite{Koeppen, Friedrichs1, Friedrichs2}.
Also in this case, there has been some diatribe about water molecules adsorbing molecularly or dissociatively~\cite{Koeppen}, but according to the latest literature~\cite{Agosta, Zhao} molecular adsorption should be preferred.

Due to its lower stability, only little information is available for rutile (001).
However, this surface is interesting because it supports the highest electrical conductivity among other rutile facets~\cite{Diebold}. 
First investigations on the annealing of this surface were performed already about forty years ago~\cite{Tait, Firment} and then retaken into account twenty years later, in $1999$, by N{\"o}rnberg et al.~\cite{Noerenberg}.
In this work, the authors were able to clearly observe a $(7\sqrt{2} \times \sqrt{2})$\emph{R}$45^\circ$ reconstruction of the surface, which was later rationalised by Fukui, Tero and Iwasawa~\cite{Fukui, Tero} in terms of a stairs-like model which spontaneously forms after annealing the sample at a temperature of about $1050$~K. 
This model suggests a surface enriched with Ti interstitial atoms, presenting a Ti$_7$O$_{12}$ stoichiometry in the primitive surface cell.
Such a reconstruction was also found to form after high-temperature epitaxial growth~\cite{Wang} as well as after pulsed-laser irradiation~\cite{Museur}.
Recently, LEED experiments proposed a similar model for (011)-faceted rutile (001) annealed at $683$~K~\cite{Ikuma}, leading to a theoretical stoichiometric reconstructed surface~\cite{Wu} which has been lately considered of main interest for the adsorption of CH$_3$OH and H$_2$O molecules~\cite{Wu2}.  

To the best of our knowledge, no study has ever addressed the relative thermodynamical stability and reactivity of these reconstruction models when interfaced with bulk water at the full DFT level.
These issues will be investigated in the present article, in which we also assess the performance of DFTB in reproducing the obtained results at a fraction of the computational cost.

\section{Simulation Details}

\subsection{Software and Methods}

In this work we employed the Vienna Ab initio Simulation Package (VASP) \cite{VASP} and the Density Functional based Tight Binding (and more) software (DFTB+)~\cite{DFTB} both for static electronic structure calculations and geometry optimizations, and for Molecular Dynamics (MD) simulations. 
The system geometries and trajectories were visualized with the Visual Molecular Dynamics (VMD) program \cite{VMD}. 

The DFT calculations were performed both with the Local Density Approximation (LDA)\cite{LDA} and the Generalized Gradient Approximation with the Perdew-Burke-Ernzerhof functional (GGA-PBE) \cite{Perdew}, in the the framework of the Projector Augmented Wave (PAW) method \cite{PAW}.
The exchange-correlation functional was extended  with the zero-damping DFT-D3 method of Grimme~\cite{Grimme} for all calculations including water molecules.
The plane-wave cutoff energy was set at $700$ eV and the Self Consistent Field (SCF) absolute errors at $10^{-4}$ eV. 
The k-points meshes corresponded to a  $3\!\times\!3\!\times\!3$ distribution for the primitive bulk supercell of the rutile crystal, rescaled according to the cell size in all other systems.
The Density of States (DOS) calculations were performed with 36 k-points for all the surfaces except for the reconstructed (001) surface, in which 18 k-points were sufficient due to the large system size, and analyzed using the vaspkit software~\cite{vaspkit}.
All MD simulations were performed sampling the cell only at the $\Gamma$-point, which was accurate enough for the purpose of the simulated reactions.

The DFTB calculations used the Tiorg Slater-Koster file set~\cite{Dolgonos, dftb_website}, where the electronic part was created using GGA-PBE, whereas the repulsive part was fitted on B3LYP energies. Because the repulsive potential in the Tiorg set has some numerical noise, we decided to smooth it by polynomial fits. These modifications are very close to the original set, but vanish much smoother at their cutoffs. The electronic part was unchanged. The set with the smoothed repulsion will be referred to as Tiorg-smooth. 
For all calculations the self-consistent charge (SCC) error was set to $10^{-6}$ electrons and the forces were allowed to relax to values below $10^{-4}$ a.u. 
The k-points meshes were chosen as in the DFT calculations.
In order to include Van der Waals interactions, the DFT-D3 method with zero damping was used, setting the parameters $s_{r,6} = 1.217$ and $s_8 = 0.722$ as provided for GGA-PBE by the Mulliken Center for Theoretical Chemistry~\cite{Bonn}. 

The MD simulations were performed using a timestep of 0.5\,fs, assigning the mass of Deuterium to the H atoms.
Thermalization of the systems was carried out by increasing the system temperature to 320~K over 1 ps by means of a Nos{\'e}-Hoover thermostat.
Afterwards, longer runs were continued in the NVE (microcanonical) ensemble.

\subsection{System cells}

\begin{figure} [ht]
\begin{center}
\includegraphics[scale=1]{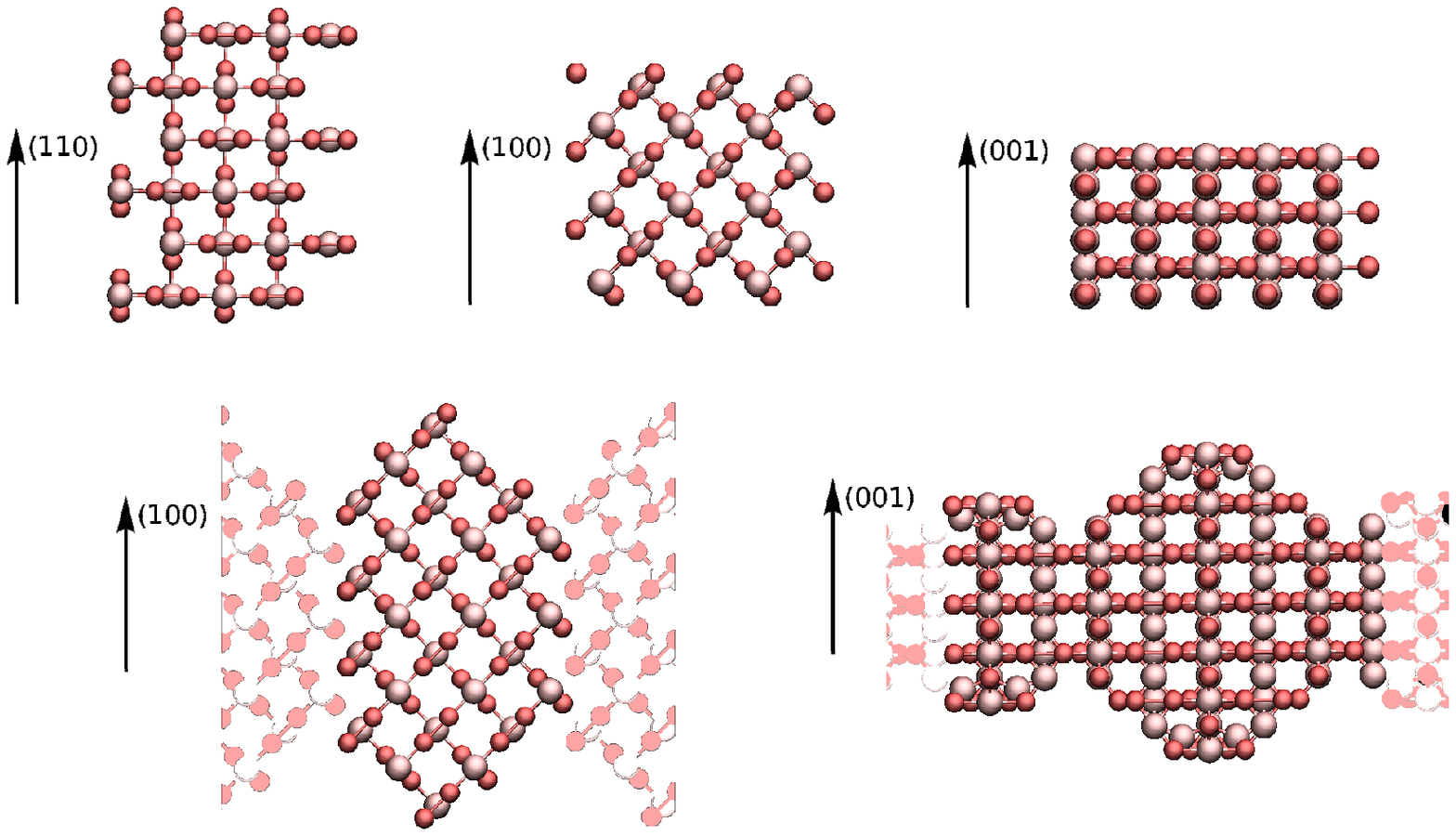}
\caption {Side view of the simulated rutile slab systems, in the top line (110), (100) and (001) and in the bottom line the two reconstructed models (100)-rec and (001)-rec. Red spheres refer to oxygen atoms, rose ones to titanium atoms. For the (100)-rec and (001)-rec slab models, the periodic images are highlighted with brighter colors.}
\label{slabs}
\end{center}
\end{figure}

All surface slabs were constructed starting from a  $3\!\times\!3\!\times\!3$ bulk rutile cell, if not otherwise stated.
Relaxation of the bulk cell until all force components were less than $10^{-4}$ au led to the equilibrium lattice parameters shown in Table~\ref{tab:lattice} in comparison with experimental and theoretical references.
\begin{table}[b]
\centering
	\caption{Lattice parameters of rutile estimated with DFT and DFTB}
	\begin{tabular}{cccccccccr}
	\toprule
	Method &a = b  & c  \\
	\midrule
	 LDA &  4.570 $\angstrom$ & 2.931 $\angstrom$ \\
	 GGA-PBE &  4.663 $\angstrom$ & 2.968 $\angstrom$ \\
     Tiorg  &  4.671 $\angstrom$ & 2.993 $\angstrom$ \\
     Tiorg-smooth  &  4.677 $\angstrom$ & 2.971 $\angstrom$ \\
	 Exp. \cite{Diebold}  &  4.587 $\angstrom$ & 2.953 $\angstrom$   \\
     HF \cite{Labat_1} &  4.575 $\angstrom$ & 2.999 $\angstrom$ \\  
     B3LYP \cite{Labat_1}  &  4.639 $\angstrom$ & 2.974 $\angstrom$ \\ 
	\bottomrule
	\end{tabular}
	\label{tab:lattice}
\end{table}
The pristine rutile (110), (100) and (001) surfaces were obtained by splitting the bulk crystal along the respective directions leading to slab thicknesses of at least six atomic planes separated by a vacuum region of at least 15\,\AA\ between the periodically repeated slabs.
During geometry relaxations only, this distance was increased up to 60\,\AA\ in the DFTB simulations without increase of the computational effort.
The non-reconstructed (110) surface was represented by a $2\sqrt{2}\!\times\!3$ primitive cell in the $\{x,y\}$ surface plane.
The non-reconstructed (100) and the reconstructed (100)-rec surfaces were represented by a $3\!\times\!3$ surface cell.
The non-reconstructed (001) surface was represented by a $3\!\times\!3$ surface cell, whereas the (001)-rec model, with stairs-like features parallel to the (011) and (0$\overline1$1) directions, was represented by a $4\sqrt{2}\!\times\! 2\sqrt{2}$ surface cell in the $\{x,y\}$ surface plane. 
All surface slabs were terminated symmetrically along the $z$ direction to avoid formation of spurious macroscopic dipoles in the cell. 
The used models are shown in Figure~\ref{slabs}.
The fully solvated systems were obtained by the iterative addition of H$_2$O molecules through the GROMACS box solvation tool~\cite{GROMACS}. 
The empty space in the unit cells was first filled at a standard water density of 1000\,g/l. 
MD simulations at a constant temperature of 320\,K were then performed for about $1.5$ ps, during which water molecule started adsorbing and created a structured solvation layer at the surfaces, thus reducing the water density between the surface slabs.
The amount of water molecules was then again increased using the box-solvation tool of GROMACS and the procedure was repeated until the bulk water density was reached far from the surfaces.
Only then the production NVE simulations were carried out, with a duration between 6 and 12 ps.

\section{Results and discussion}

In the first part of this section, we will present static calculations with geometry relaxation of all pristine and reconstructed surfaces, both regarding their energy and their electronic structures.
In the second part, the different reactivities of the surfaces towards water molecules, with particular attention paid to the reconstructed surfaces in contact with bulk liquid water, will be presented.

\subsection{Surface structures and energies}

For stoichiometric systems, the surface energy of a slab model can be calculated as~ \cite{Perron, Kiejna, Labat, Song, Reuter, Gonzales}
\begin{equation}
    \gamma_{\text{Surf}} = \frac{E_\text{Slab} - N_{\text{Ti}} E_\text{Bulk}}{2\text{A}} \text{ ,}
    \label{eq:surface_energy}
\end{equation}
where $E_\text{Slab}$ is the total energy of the slab model containing $N_{\text{Ti}}$ atoms and $E_\text{Bulk}$ is the energy of a bulk TiO$_2$ unit.
The exposed surface area is twice the area $A$ of the $\{x,y\}$ plane in the surface primitive cell.

\begin{figure}[ht]
\begin{center}
\includegraphics[scale = 1]{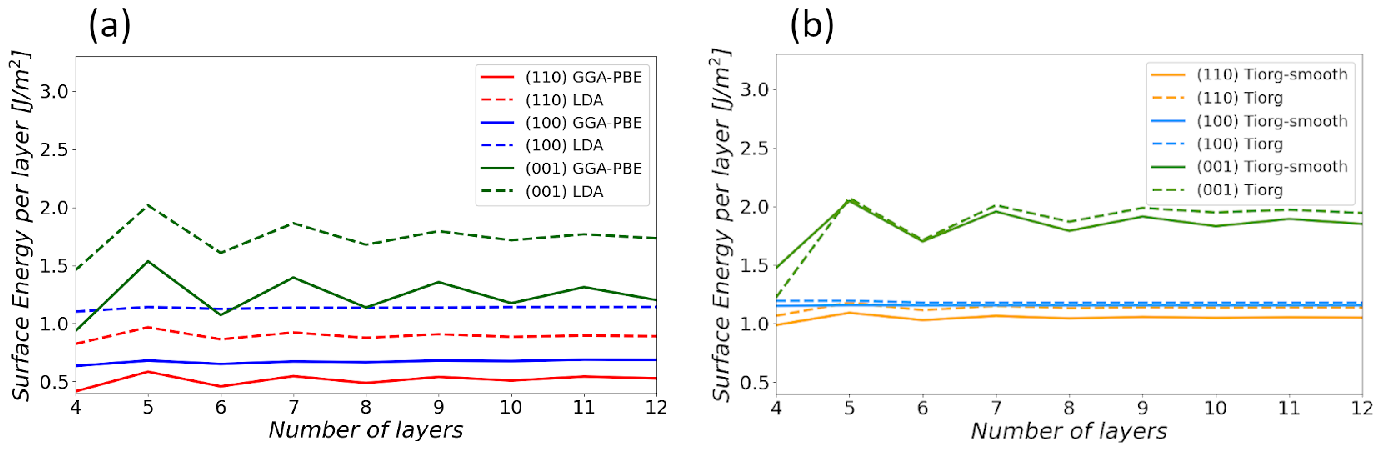}
\caption {The surface energy $\gamma_{\text{surf}}$ with respect to the number of layers for the respective rutile slab models is plotted for the (a) DFT and (b) DFTB approaches. The surfaces are indicated by different colors, namely  (110) in red, (100) in blue and (001) in green with lighter nuances for the DFTB approach.}
\label{pristine_geo_curves}
\end{center}
\end{figure}

As expected, $\gamma_{\text{Surf}}$ converges only slowly and with a typical oscillatory behaviour with respect to the number of atomic layers in the slab, as shown in Figure~\ref{pristine_geo_curves}.
Due to the different truncation geometries along the $z$ axis (see Figure~\ref{slabs}), this behaviour is most pronounced for the strongly asymmetric (001) surface and barely appreciable for the perfectly symmetric (100) surface.
The values for 12 atomic layers, reported in Table~\ref{tab:convergence}, are in very good agreement with previous theoretical works both for DFT \cite{Perron, Kiejna, Labat, Song} and DFTB \cite{Dolgonos, Fox}. 
The correct order of stability, $(110) < (100) < (001)$, is adequately described in all the different methods. 
\begin{table}[ht]
\centering
	\caption{Converged values of the surface energy $\gamma_{\text{Surf}}$ in $\frac{J}{m^2}$ for rutile-(001), (100) and (110) for the different approaches}
	\begin{tabular}{cccccccccr}
	\toprule
	&LDA  & GGA-PBE & Tiorg & Tiorg-smooth & HF \cite{Labat} & B3LYP \cite{Labat} \\
	\midrule
	(001) &  1.75  & 1.26  & 1.94  & 1.87 & 2.08 & 1.45 \\
	(100) &  1.15  & 0.69  & 1.18  & 1.16 & 1.13 & 0.70  \\
	(110) &  0.89  & 0.53  & 1.14  & 1.05 & 0.92 & 0.46  \\
     
	\bottomrule
	\end{tabular}
	\label{tab:convergence}
\end{table}
However, the surface energy values differ significantly depending on the used method. 
Whereas the GGA-PBE values agree well with the most precise B3LYP literature values~\cite{Labat}, the LDA values are overestimated, similarly to the values predicted by pure HF~\cite{Labat}.
Notably, both DFTB parameter sets also overestimate $\gamma_{surf}$ in all cases, giving results that are very similar to LDA, despite the fact that the parametrization was performed with GGA-PBE as a reference for the attractive interactions.
Among the two choices of parameter sets, the Tiorg-smooth performs slightly better, especially regarding the difference between the surface energy of the (110) and the (100) facets. 

The reconstructed (100)-rec surface presents a repeated roof-like structure consisting of sharp edges and grooves faceted along the (110) and the chemically equivalent ($1\overline{1}$0) directions. 
While the Ti atoms at the bottom of the groove are six-fold coordinated by O atoms, as in the bulk, the edges consist of rows of five-fold coordinated Ti atoms bound to bridging O atoms.
During geometry relaxation, these O atoms move from their bulk position to a symmetric position at the edge top and the edge-angle increases from 45$^{\circ}$ to about 65$^{\circ}$ as a consequence.
Overall, however, the facets maintain almost perfectly the chemical features of the (110) surface.
Since the resulting exposed facet area is $\sqrt{2}$ larger than the nominal area of the surface plane (to which $\gamma_{surf}$ is referred to), we can expect the surface energy of (100)-rec to be approximately $\sqrt{2}$ times larger than the surface energy of the pristine (110) surface.
This is highlighted in Table~\ref{tab:roof}, where deviations from the ideal factor are indicative of an overall slightly reduced chemical reactivity of the surface with respect to the (110) plane.
It has to be noted, however, that the exposed edges may still act as reaction hotspots when the surface is in contact with a humid environment.
In fact, $\gamma_{surf}$ is {\em higher} for (100)-rec than the pristine (100) facet (0.73 vs 0.69~J/m$^2$). Experimentally, the roof-like reconstruction occurs at high temperature probably due to favourable vibrational entropy.
Regarding the different approximation levels used to compute $\gamma_{surf}$, GGA-PBE presents an energy ratio of 1.38, and thus a lower reactivity than LDA, for which the energy ratio is exactly $\sqrt{2}$.
The Tiorg DFTB set strongly underestimates the energy ratio, while the Tiorg-smooth is in good agreement with the GGA-PBE value.

\begin{table}[ht]
\centering
	\caption{Surface energy values $\gamma_{\text{Surf}}$ in $\frac{J}{m^2}$ of the (100)-rec slab model for the DFT and DFTB approaches, listed as absolute values and in relation to the pristine (110) surface }
	\begin{tabular}{cccccccccr}
	\toprule
	&$\gamma_{\text{Surf}}$  & \text{Ratio w.r.t.} (110) \\
	\midrule
	 LDA &  1.26  & 1.41 $\sim 1.00 \sqrt{2}$    \\
	 GGA-PBE &  0.73  & 1.38 $\sim 0.97\sqrt{2}$   \\
	 Tiorg &  1.47 & 1.29 $\sim 0.91\sqrt{2}$  \\
	 Tiorg-smooth &  1.42  & 1.35 $\sim 0.95\sqrt{2}$  \\
     \bottomrule
	\end{tabular}
	\label{tab:roof}
\end{table}

\begin{figure}[ht]
\begin{center}
\includegraphics[scale = 1]{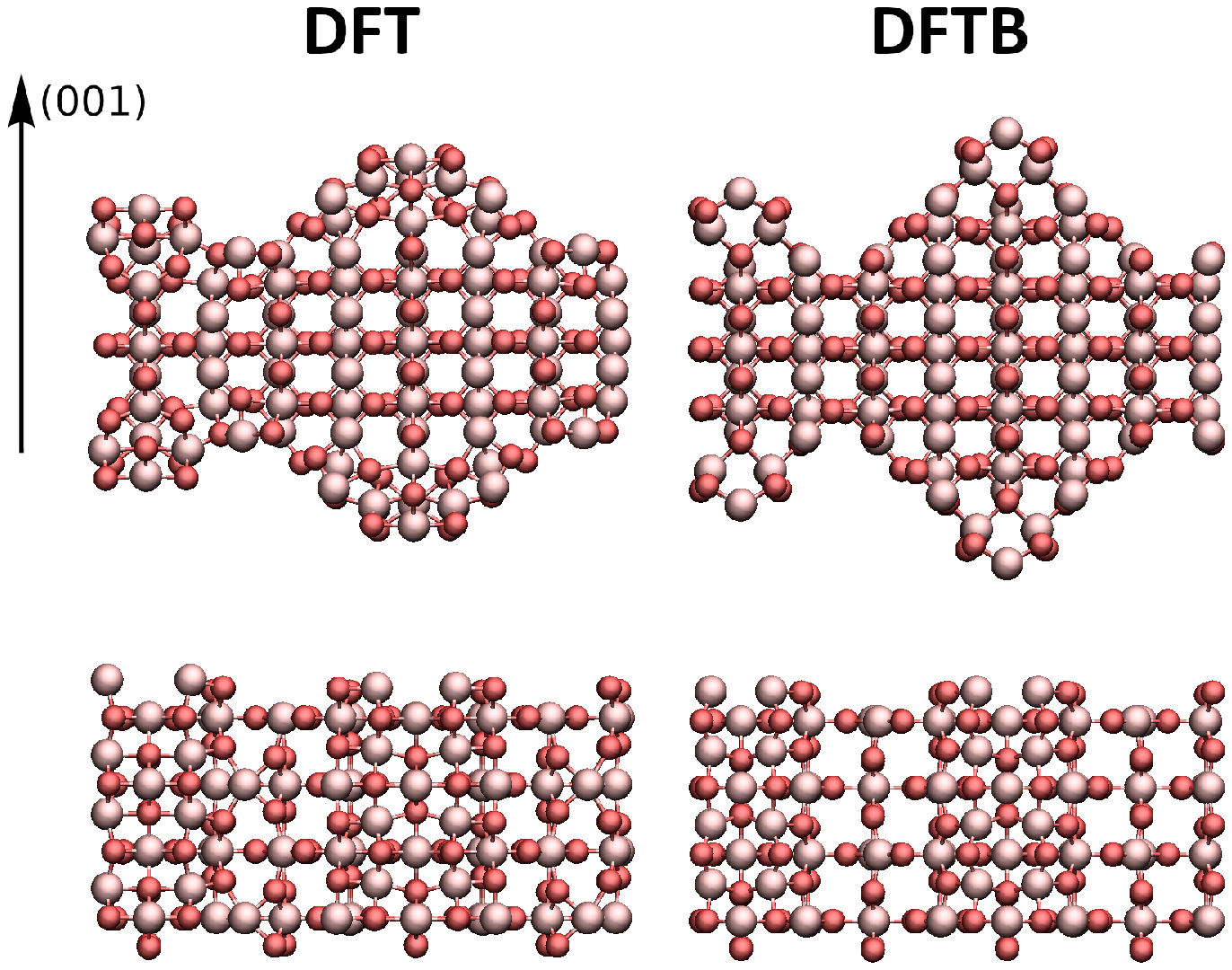}
\caption {Optimized geometries of the (001)-rec model are plotted for the DFT and DFTB approaches as side and top view. Only GGA-PBE (DFT) and Tiorg-smooth (DFTB) configurations are illustrated, as they are representative for LDA (DFT) and Tiorg (DFTB) as well. Oxygen atoms are shown as red spheres and titanium atoms as pink ones.}
\label{Stairs_Fil_Verena}
\end{center}
\end{figure}

In the case of the non-stoichiometric (001)-rec surface, geometry optimization leads to two different final configurations at the DFT and DFTB levels as shown in Fig.~\ref{Stairs_Fil_Verena}, with no differences between LDA and GGA or betweeen Tiorg and Tiorg-smooth, respectively.
For DFT, the sharp boundaries between the islands and the valleys (cf. Fig.~\ref{slabs}) become smoother as consequence of both the O atoms and the Ti atoms increasing their respective coordination shell.
Notable is the strong relaxation around the Ti interstitials of the higher island, where the orginally five-fold coordinated O atoms break their bonds with the Ti atoms right underneath.
In the resulting structure, all O atoms are at least three-fold coordinated and all Ti atoms are at least five-fold coordinated, although with strong distortion of their coordination shells.
This gives the outermost surface layer a rather amorphous character, which is not uncommon for Ti oxides~\cite{Schneider_2010}.
The final DFT structure resembles the model proposed by Ikuma et al.~\cite{Ikuma} for rutile-(001) after annealing at 683\,K, in particular regarding the flattening of the sharp groove bottom and the presence of kinks along the (011) directions.   

The relaxed surface structure is evidently different for DFTB, where the interstitial Ti rows moves from a sub-surface position to an ad-atom position on top of both islands, and remains coordinated by  only two O atoms.
Unfortunately, the available experimental information does not allow us to judge whether the DFT or the DFTB predicted structure is correct.
Both configurations are compatible with the LEED symmetry and with the periodicity of the features observed in top-view with STM and AFM experiments.
Hence, we carried out an analysis of the surface energies, in particular attempting a prediction of the expected transition temperature between the pristine (001) termination and the (001)-rec reconstruction. 

The evaluation of the surface energy for this reconstructed case shall be treated differently than before, since the presence of Ti interstitials leads to a surface stoichiometry equivalent to Ti$_7$O$_{12}$.
The Ti excess can be taken into account by an {\em ab initio} thermodynamic ansatz, in which the surface is thought to be in equilibrium with an oxygen atmosphere at a certain pressure $p$ and temperature $T$, in which the chemical potential of the O atoms is
\begin{equation}
    \mu_{\text{O}} (T,p) = \Tilde{\mu}_{\text{O}}(T, p_0) + \frac{1}{2}k_B T \ln{\frac{p}{p_0}}  \text{.}
\end{equation}
Here, $\Tilde{\mu}_{\text{O}}(T, p_0)$ is a reference O chemical potential at the standard pressure $p_0 = 1$\,atm, as tabulated in the JANAF thermochemical tables~\cite{JANAF}, and $k_B$ is the Boltzmann constant.
Neglecting the entropy differences between the bulk and the surface system, which introduces an error of the order of only a few meV per cell for oxide systems of similar size~\cite{Reuter}, the surface energy of the non-stoichiometric surface system including $N_{Ti}$ Ti atoms and $N_O$ O atoms can be expressed as~\cite{Agosta, Zhao, Gonzales}:
\begin{equation}
     \gamma_{\text{Surf}}(T,p) = \frac{ E_{\text{Slab}} - N_{\text{Ti}} E_{\text{Bulk}} + (2N_{\text{Ti}} -  
             N_{\text{O}})\mu_{\text{O}}(T,p)}{2\text{A}} \text{ .}
    \label{eq:surface_energy_modified}   
\end{equation}

\begin{figure}[ht]
\begin{center}
\includegraphics[scale = 1]{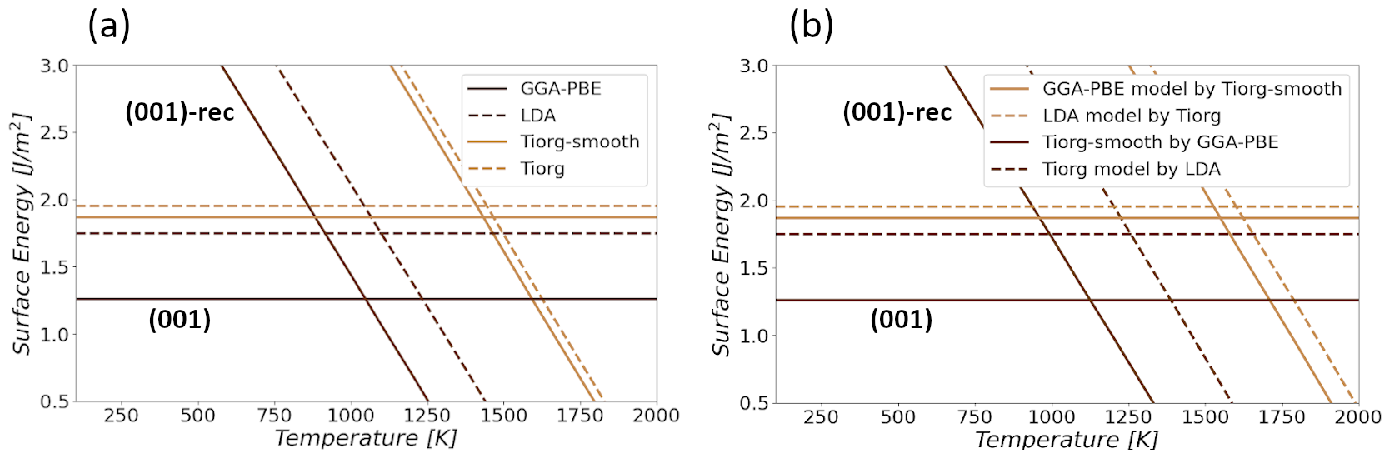}
\caption {(a) The surface energy $\gamma_{\text{Surf}}$ of the reconstructed (001)-rec slab model, as indicated in Figure \ref{Stairs_Fil_Verena} and the pristine (001) slab model against the temperature is plotted for the DFT approach with GGA-PBE (brown, solid line) and LDA (brown, dashed line) functionals. Results for the DFTB approach are illustrated for the Tiorg-smooth set (beige, solid line) and the Tiorg set (beige, dashed line). (b) The dependence of the surface energy $\gamma_{\text{Surf}}$ on the temperature was calculated here using the optimized (001)-rec structure of the respective other methodological approach: GGA-PBE/ DFT calculation with the Tiorg-smooth/ DFTB optimized structure (brown/solid line), LDA/ DFT with Tiorg/ DFTB (brown/ dashed line), Tiorg-smooth/ DFTB with the GGA-PBE/ DFT (beige/ solid line) and  Tiorg/ DFTB with the LDA/ DFT optimized structure (beige/ dashed line). All formation temperatures are shifted to higher values in comparision to the approaches in (a). }
\label{001-rec_geo_curves}
\end{center}
\end{figure}

The resulting $\gamma_{\text{Surf}}(T)$ values at a pressure $p = 5\cdot 10^{-8}$\,atm, as set in the experiments by Tero et al.~\cite{Tero}, are reported in Fig. \ref{001-rec_geo_curves}(a) for the DFT and DFTB structures shown in Fig.~\ref{Stairs_Fil_Verena}.
The constant lines correspond to the surface energies obtained with the various formalisms (GGA, LDA, Tiorg and Tiorg-smooth) for the stoichiometric, pristine (001) surface.
The intersections with the $\gamma_{\text{Surf}}(T)$ curves relative to the (001)-rec surface represent the predicted transition temperatures at which spontaneous surface reconstruction occurs.
Experimentally, the reconstruction was observed to take place right above 1050\,K in the work of Tero et al.~\cite{Tero}, and at 1027\,K in the work of N{\"o}renberg et al.~\cite{Noerenberg}
These values agree remarkably well with the GGA-PBE prediction of 1048\,K (Table~\ref{tab:temps}), while the LDA prediction of 1098\,K is slightly larger.
Severely overestimated, instead, are the values predicted by DFTB, which lie above 1400\,K independently of the used set of parameters.
In order to check whether the obtained structures at the two approximation levels may be trapped in local minima of their respective potential energy surface, the structure obtained with GGA-PBE was further relaxed using DFTB, and the structure obtained by DFTB was relaxed using DFT.
In both cases, relaxation did not lead to major structural rearrangements, but the obtained total energies were {\em higher} than with the original models.
This leads to a shift of the transition temperatures for the reconstruction towards larger values, as reported in 
Figure~\ref{001-rec_geo_curves}(b) and in Table~\ref{tab:temps}.

It has thus to be concluded that DFT, being able to predict very well the experimentally observed transition temperature (especially at the GGA-PBE level), most probably also delivers a correct atomic configuration of the reconstructed surface structure.
On the other hand, both DFTB parametrizations are {\em not} able to reproduce the correct structural and energy features of this surface.
Since the attractive part of the DFTB interaction was parametrized on GGA-PBE calculations, most probably the source of discrepancy lies in the two-body repulsive interaction, which appears not to be transferable to non-stoichiometric situations.
In particular, DFTB tends to underestimate the penalty associated with the presence of strongly undercoordinated Ti atoms (the apex Ti row in the reconstructed islands visible in Fig.~\ref{Stairs_Fil_Verena}), and to overestimate the penalty associated with the formation of strongly distorted coordination shells around partially reduced Ti atoms, which is a structural feature clearly observed in DFT.

\begin{table}[ht]
\centering
	\caption{Calculated formation temperatures for the reconstruction on rutile-(001) evaluated from Figure ~\ref{001-rec_geo_curves}.}
	\begin{tabular}{cccccccccr}
	\toprule
	&\text{Standard relaxation}  & \text{Reversed relaxation} \\
	\midrule
	 GGA-PBE &  1048 K & 1131 K     \\
	 LDA &  1098 K  & 1257 K    \\
	 Tiorg &  1445 K  & 1601 K   \\
	 Tiorg-smooth &  1433 K  & 1549 K  \\ 
     \bottomrule
	\end{tabular}
	\label{tab:temps}
\end{table}

\subsection{Density of states}

\begin{figure}[H]
\begin{center}
\includegraphics[scale = 1]{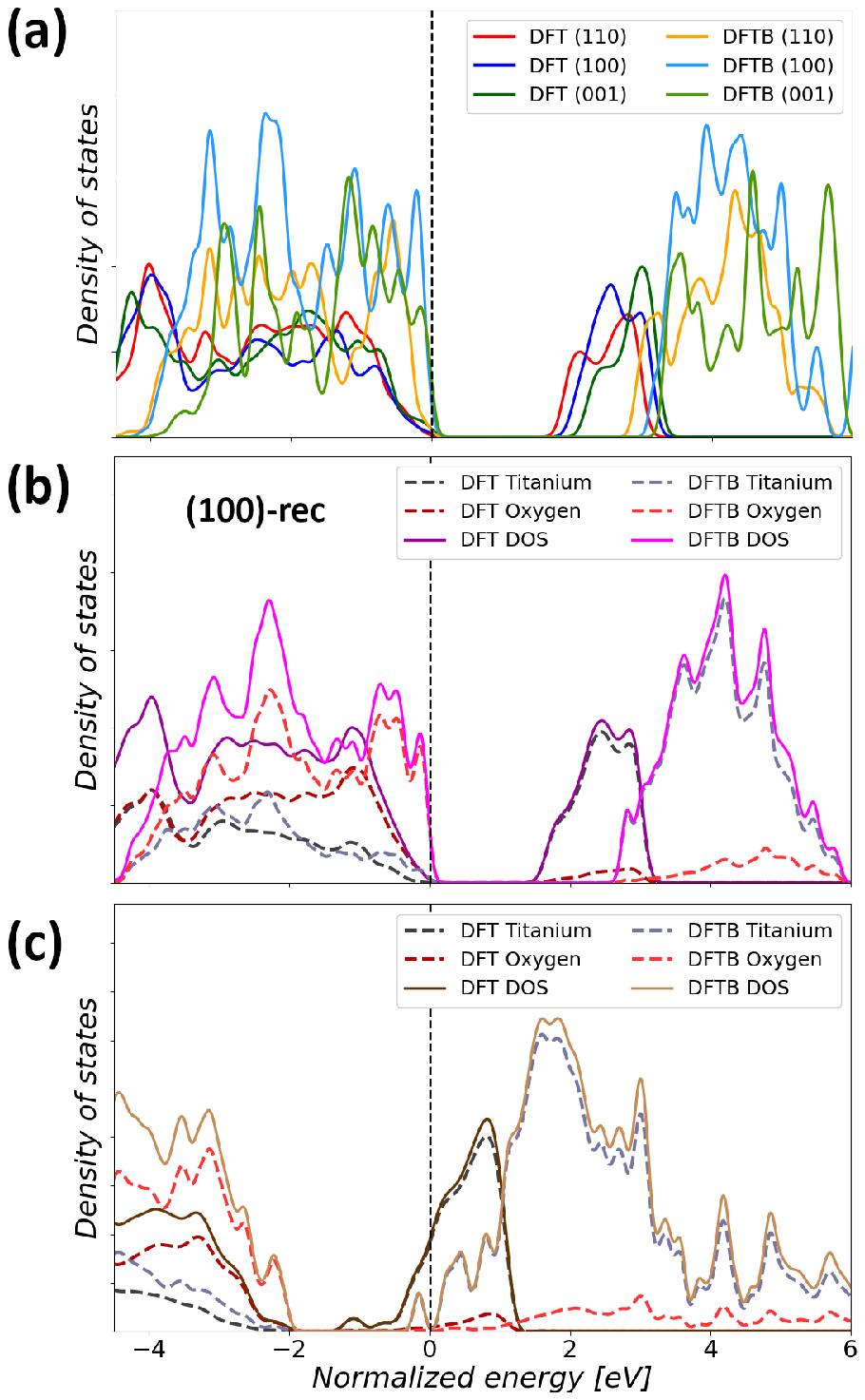}
\caption {Top view: (a) Density of States (DOS) provided by the surface atoms of rutile-(110), (100) and (001). Bottom view: Projected Density of States (PDOS) provided by the surface atoms of (100)-rec (b) and (001)-rec (c). all shifted at the \emph{valence band edge} (VBE). }
\label{surface_PDOS}
\end{center}
\end{figure}

Before starting the investigation of the reaction of water with the surface models presented in the previous section, it is important to gain information about their electronic structure.
To this aim, we perform an analysis of the partial density of states (PDOS), limiting the study to the GGA-PBE functional in DFT and the Tiorg-smooth parameter set in DFTB.
The PDOS of all atoms belonging to the topmost surface layers of the three pristine surfaces are shown in Figure~\ref{surface_PDOS}(a).
The most notable difference between the two formalisms is that DFTB predicts band gaps about 1.1\,eV larger than DFT (Table~\ref{tab:all_PDOS}).
While underestimation of the band gap is a classic problem of DFT in the local approximation due to the self-interaction error~\cite{Perdew2, Deak}, DFTB corrects for this effect by adjusting the compression radii in the repulsive part of the Slater-Koster  files~\cite{Dolgonos}.
Another important difference is the very steep increase of the DOS at the valence-band edge obtained with DFTB, whereas DFT predicts a much smoother band edge.

\begin{table}[H]
    \centering
    \begin{tabular}{cccc}
	\toprule
	    &\text{DFT}  & \text{DFTB} & \\
	    \text{Surface} &\text{(GGA-PBE)}  & \text{(Tiorg-smooth)} & \text{Difference} \\
	    \hline
	    (110)  &  1.75  & 2.87 & 1.12   \\
	    (100)  &  1.94  & 3.14 & 1.20    \\
	    (001)  &  2.18  & 3.26 & 1.08 \\ 
	    \enspace (100)-rec. &  1.59  & 2.78 & 1.19   \\
	    \enspace (001)-rec &  0.00  & 0.00 & 0.00   \\
        \bottomrule
	    \end{tabular}
    
    \caption{Band gap values in eV of both the pristine and reconstructed surfaces.}
    \label{tab:all_PDOS}
\end{table}
This difference is attributable to the larger electron density on the O atoms, which contribute for the largest part to the valence-band edge, whereas the Ti atoms contribute mostly to the conduction-band edge above the band gap.
This is exemplified for the case of the (100)-rec surface model in Figure~\ref{surface_PDOS}(b).
This surface does present a slightly smaller band gap than the pristine (110) facets which terminate the surface (cf. Figure~\ref{slabs}).
Otherwise, the same differences between the DFT and DFTB features as for the pristine facets are observed, in particular regarding the steeper onset of the valence-band edge.

The electronic structure of the (001)-rec model is quite different, because the excess Ti interstitials confer to the surface a metallic character, as shown in Figure~\ref{surface_PDOS}(c).
Here, the conduction-band edge shifts below the Fermi level, to a larger and more evident extent in DFT than DFTB.
In fact, while SCC-DFTB is in principle able to describe correctly metallic systems, both the Tiorg and Tiorg-smooth sets predicted the wrong structure and energetics for this particular reconstruction. However, it is not easy to determine the reason of this mismatch, as it could rather be the combination of three main factors. First, the difference in the geometry already results in different electronic structures. Secondly, the gap sizes in the two methods are not the same either and, lastly, the Slater-Koster parametrization should be specific for the metallic or the isolator case.

\subsection{Interaction with water}

We turn now to the investigation of water interacting with the surfaces both as single molecule and as part of a bulk liquid.
Particular emphasis will be given to the ability of the Tiorg-smooth DFTB parameter set in reproducing the GGA-PBE DFT results.
In both cases, the calculations are corrected with the DFT-D3 dispersion term of Grimme \cite{Grimme}.

\subsubsection{Static calculations at low water coverage}

\begin{figure}[H]
\centering
\includegraphics[scale = 1]{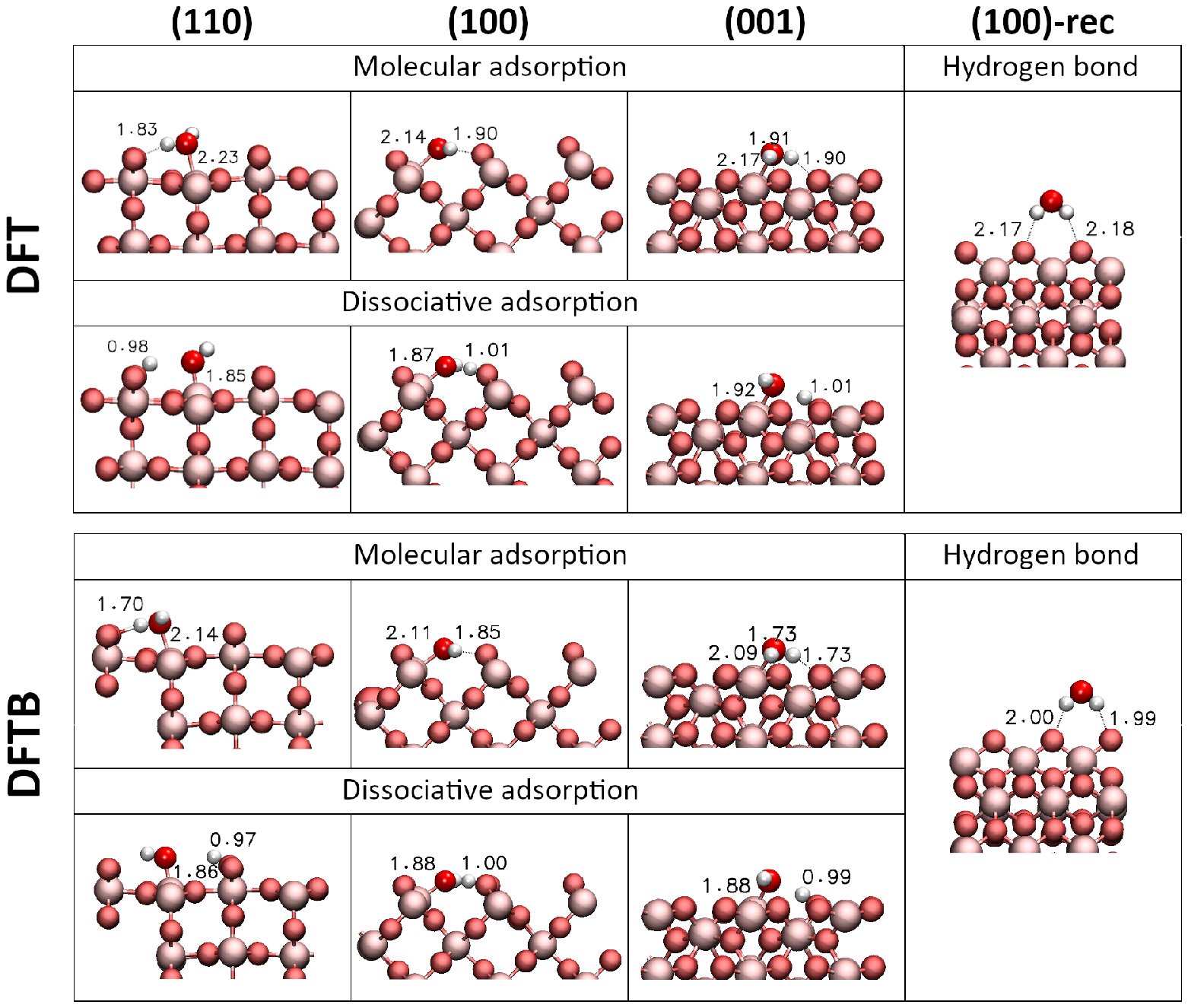}
\caption{Final geometries of the adsorption of a single water molecule on the pristine (110), (100), (001) and reconstructed (100) surface slab models are illustrated as ball and stick model with metallic red and pink colors for the surface oxygen and titanium atoms. Water is colored in pure red and white balls for oxygen and hydrogen atoms. The DFT results refer to GGA-PBE exchange-correlation functional calculations and the DFTB results are performed with the Tiorg-smooth set of parameters. For each pristine surface, the molecular and dissociative adsorption mode has been analysed. Relevant distances are labeled with their equilibrium values. }
\label{static_adsorptions}
\end{figure}

As a first step, single H$_2$O molecules were adsorbed on the pristine (110), (100) and (001), and the reconstructed (100)-rec surfaces.
Two scenarios were taken into account: molecular adsorption and dissociative adsorption, in line with much of the known literature~\cite{Agosta, Zhao, Gonzales}. 
The final, optimized geometries are shown in Figure~\ref{static_adsorptions} and the correspondent adsorption energies per molecule
$\Delta E^{\text{mol}}_{\text{ads}} = \frac{1}{N_{\text{H}_2\text{O}}}\big(E_{\text{slab+H$_2$O}} - E_{\text{slab}} - N_{\text{H}_2\text{O}} \cdot E_{\text{H}_2\text{O}}\big) $ 
are reported in Table~\ref{tab:par}.

In the case of molecular adsorption, stable geometries were found in DFT on all of the three pristine surfaces with Ti(surf)-O(water) distances slightly larger than the usual Ti(bulk)-O(bulk) value of 1.95\,\AA~\cite{Diebold}. 
The configurations were further stabilized by hydrogen bonds with adjacent O surface atoms. 
Comparable geometries were found in the DFTB systems with systematically slightly shorter hydrogen bond distances, but in overall good agreement with the DFT references. 
In the case of dissociative adsorption, the Ti(surf)-O(surf) of the covalently bonded OH group was even shorter than the Ti(bulk)-O(bulk) value. 
The corresponding proton transferred to an adjacent oxygen surface atom formed a hydrogen bond with the adsorbed hydroxyl group. 
This configuration was found on all of the three pristine surface models. 
Also in this case, DFTB reproduced the geometries in very good agreement with DFT. 
The additional adsorption configuration on the (100)-rec structure, which included exclusively hydrogen bonds at the tip of the edges, is again stabilized with the same geometry in both DFT and DFTB. 
In this case, though, we found the H(water)-O(surf) distances in DFTB to be $0.2$ $\angstrom$ shorter than in DFT.

\begin{table}[ht]
\centering
	\caption{Adsorption energies $\Delta E^{\text{mol}}_{\text{ads}}$ (eV) of water with respect to different adsorption modes as illustrated in Figure \ref{static_adsorptions} on rutile-(110), (100) and (001) as well as (100)-rec are listed. The reported values rely either on single molecule water adsorption modes or water embedded in a monolayer configuration.  The most stable configuration for each surface model is highlighted in bold.}
	\begin{tabular}{ccccccccccr}
	\toprule
	Coverage & Surface & Ads. mode  & DFT   & DFTB
	\\
	& &  & \text{(GGA-PBE)}  & \text{(Tiorg-smooth)}
	\\\midrule
    1 H$_2$O &&&& \\
	\midrule
	& (110) &  Mol.   & \textbf{-0.99}    & -1.00     \\
	 \midrule
	 && Diss.   & -0.82  &   \textbf{-1.29}     \\
	 \midrule
	 &	(100) &  Mol.  &  \textbf{-1.21}  &        \textbf{-1.28}    \\
	 \midrule
	 &&  Diss.      &  -0.97  &       -0.94    \\
	 \midrule
	 &	(001) &  Mol.       &  \textbf{-1.23}       &  -1.46    \\
	 \midrule
	 &&  Diss.      &  -1.09       &  \textbf{-1.61}    \\
	 \midrule
	 & (100)-rec & Hydr.  &  -0.21    &  -0.38    \\
     \midrule
     1 ML &&&& \\
	\midrule
	 & (110) & Mol. &   -1.11  & -1.52   \\
	 \midrule
	  && Diss. &  -0.84   &  \textbf{-1.86}  \\
	 \midrule
	 && Mix.  &  \textbf{-1.25}  &   -1.56    \\
	 \bottomrule
	\end{tabular}
	\label{tab:par}
\end{table}

Although DFTB reproduces with very good accuracy the structural details of both investigated adsorption modes, inconsistencies emerge when the energetics of the systems are considered.
At the DFT level, molecular adsorption is more favourable on all pristine surfaces (Table \ref{tab:par}).
Only when the water coverage is increased to 1 monolayer, a mixed molecular/dissociative adsorption mode becomes the most stable for rutile-(110), which is in good agreement with the available literature~\cite{Gonzales, Harris, Kamisaka, Luschtinetz, Agosta}.
DFTB, instead, favours dissociative adsorption in all cases apart from the pristine (100) surface.
Interstingly, the $\Delta E^{\text{mol}}_{\text{ads}}$ values computed with DFTB for molecularly adsorbed water agree well with the corresponding DFT references, whereas the values for dissociatively adsorbed water are strongly overestimated.

We believe that this inconsistency originates from two factors.
First, the strength of H-bonds is overestimated by DFTB~\cite{Goyal}.
Therefore, configurations with larger amount of in-surface H-bonds, such as the ones rich in terminal OH groups, become over-stabilized.
In fact, in the only case where molecular adsorption is favored by DFTB (the pristine (100) surface), the molecule is involved in two hydrogen bonds at once with neighbor O atoms (see Figure~\ref{static_adsorptions}).
Second, the O atoms of the surface, including those of water adsorbates that saturate Ti dangling bonds, have larger electronegativity in comparison with DFT.
This was evident by the steeper valence-band edges and larger electron density in the PDOS stemming from O atoms (see Figure~\ref{surface_PDOS}).
As a result of the stronger electron donation from Ti into the O atom of the adsorbing water molecule, the O-H bonds becomes more polarized than in the DFT case, and thus more prone to proton transfer to a neighboring O acceptor.

\subsubsection{Molecular Dynamics in the presence of liquid water}

\begin{figure}[H]
\centering
\includegraphics[scale = 1]{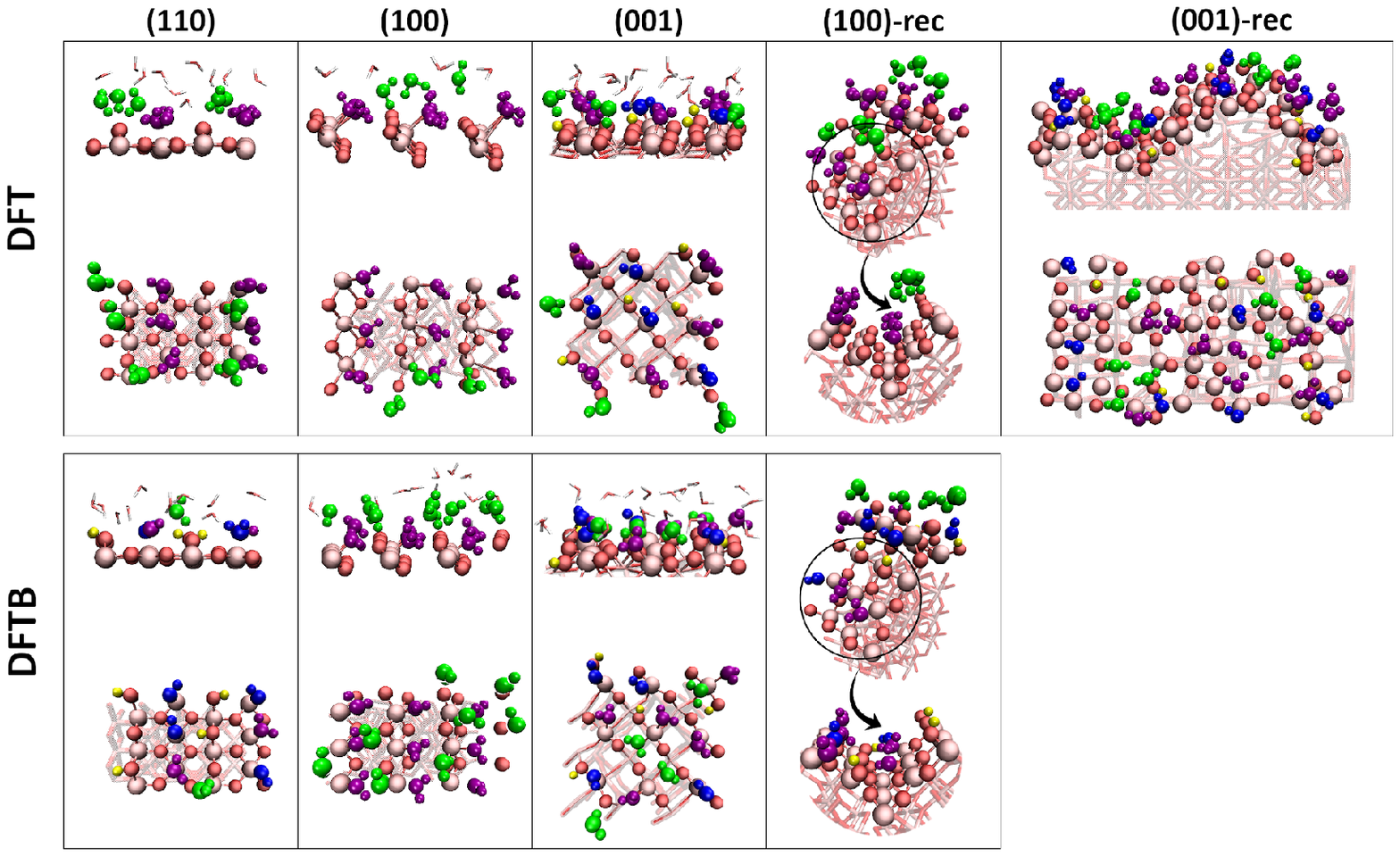}
\caption{Representative snapshots of bulk water-titania surface systems after Molecular Dynamics simulations at 300 K. The surface slab model is colored in metallic red and pink balls for the oxygen and titanium surface atoms. The first layer of water is highlighted as ball and stick model with the following coloring code: molecular adsorption (purple), hydroxylation (blue) and protonation (yellow) upon dissociation of water and hydrogen bonded water (green). Bulk water and bulk slab models are illustrated as transparent lines. }
\label{all_h2o}
\end{figure}

In addition to the static calculations at low water coverage presented above, MD simulations of bulk water in contact with the different surface models were performed. 
In this way, we studied which surface terminations are the most favourable in the presence of a fully developed water hydrogen-bond network. 
The final equilibrium configurations after thermalization of the system and production MD runs lasting between 6 and 12\,ps are shown in Figure~\ref{all_h2o}.

At the DFT level, only on the pristine (001) surface four out of nine chemisorbing water molecules dissociated within 6\,ps of MD.
In all other systems, water adsorbed molecularly and did not dissociate for the entire duration of the simulations up to more than 10\,ps.
This was true even for the pristine (110) surface, where a mixed adsorption mode is stable at the coverage of 1 ML.
However, starting from this configuration in contact with liquid water led to complete recombination of the originally dissociated water molecules within 3\,ps.
Typically, the adsorbed water molecules arranged in a more upright position with respect to the minimum configurations obtained with single water molecules, in order to engage in hydrogen bonds with the upper liquid water layers.
In-surface hydrogen bonds were also present in some cases, notably in the deep grooves of the (100)-rec surface model, where water molecules adsorbed on five-fold coordinated Ti atoms on one slope, forming one or even two hydrogen bonds at once with bridging O atoms of the opposite slope (Figure~\ref{all_h2o}).

At the DFTB level, again a much stronger tendency to adsorb dissociatively was observed.
Mixed layers including both dissociated and non-dissociated water molecules were obtained on all systems, with the exception of the pristine (100) surface, as in the static calculations.
In agreement with the behaviour observed in DFT, the adsorbed OH and OH$_2$ groups readily engaged in hydrogen bonds with the layers of liquid water above the surface, although  in-surface hydrogen bonds were also observed.

\subsubsection{Water reactions with the reconstructed (001) surface}

Particular attention was given to the reaction of the non-stoichiometric (001)-rec surface with water, because of two reasons.
First, this system may be considered representative of severe reconstructions that take place either after treatment of low-Miller-index single-crystal factes at high temperature or as the result of relaxation of high-Miller-index facets, present for instance in TiO$_2$ powder materials~\cite{Laube_TiO2_nanoparticles}.
Second, the presence of excess Ti interstitials in the system is intriguing, given the overall reduced character of the surface, which could thus behave qualitatively similar to surfaces with Ti$^{3+}$ defects that are known to form after interaction of titania with UV light~\cite{Shultz_1995, Fujishima_2000, Andrew_2008}.

Our calculations were performed only at the GGA-PBE DFT level, since DFTB was not able to predict the correct structure and energetics of this system (see above).
Due to the large system size, we decreased the cutoff energy to 400\,eV and increased the simulation temperature to 350\,K to speed up the water diffusion in the MD production run, which lasted 6 ps after thermalization.
Taking into account both the top and bottom surfaces of the slab models, a total of 36 water molecules formed direct bonds with Ti atoms of the surface.
Out of these, 24 remained adsorbed molecularly and 12 dissociated to form terminal OH groups, saturating all dangling bonds of the fivefold and four-fold coordinated Ti atoms of the dry surface.
Furthermore, other 21 molecules became incorporated into the first surface hydration layers, physisorbed via hydrogen bonds to surface O atoms.
All these adsorption modes were already observed in the static calculations (see Fig.~\ref{static_adsorptions}).

\begin{figure}[H]
\centering
\includegraphics[scale = 1]{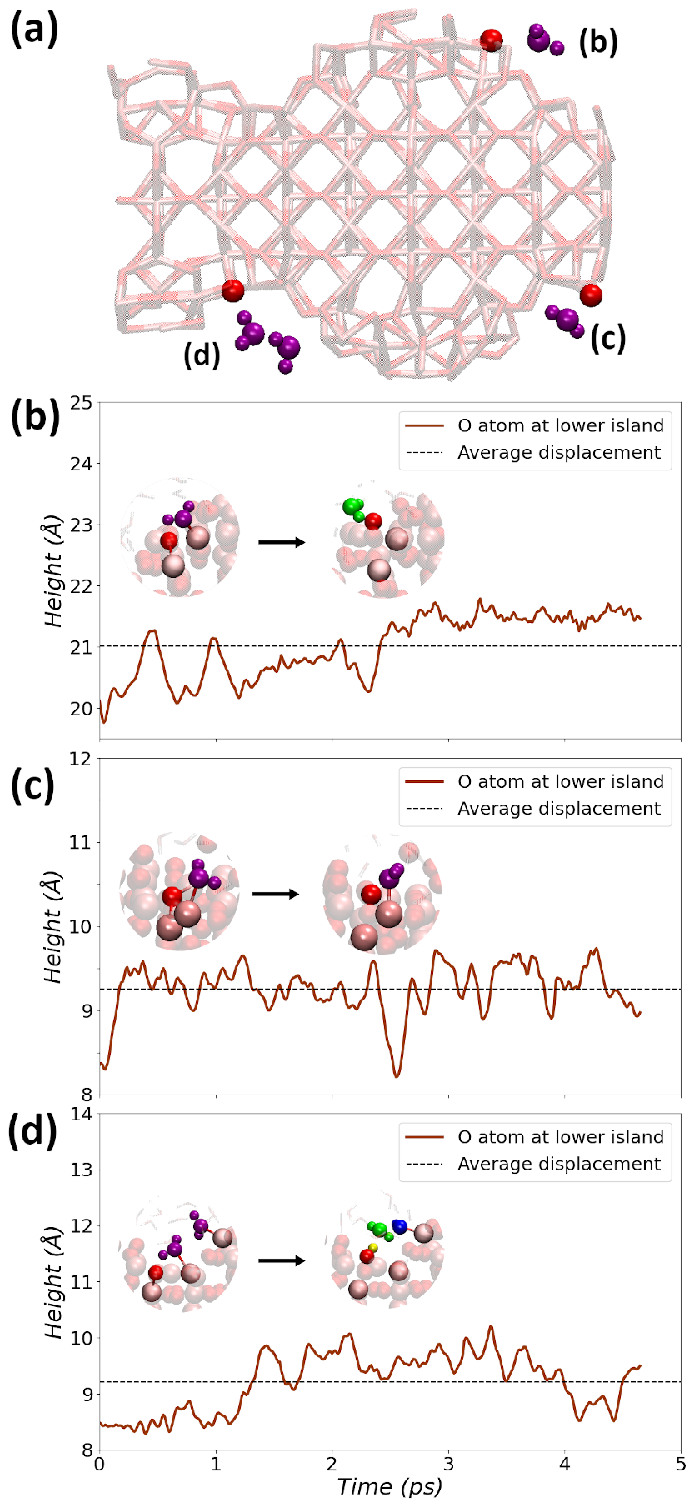}
\caption{(a) The (001)-rec surface model is illustrated in transparent side view. Three reactive oxygen surface atoms from either the island (b) or valley positions (c, d) and the corresponding adsorbed water molecules are highlighted as opaque ball and stick models. The adsorption configuration mode of the water molecule is represented with the color scheme used in Figure \ref{all_h2o}. The displacements of these highlighted oxygen atoms are plotted in (b), (c) and (d), the z coordinate of which is illustrated at each time step as brown solid line. The mean height is plotted as black dashed line. All panels show a clear dragging of the respective oxygen atom towards the water network and the involvement of neighbored adsorbed water molecules. The inserts indicate the starting and representative final configuration of each depicted oxygen surface atom. (b) The surface oxygen atom is lifted up by 2 $\angstrom$ and takes over the position of an adsorbed water molecule. (c) The surface oxygen atom is lifted up by 1 $\angstrom$ and loses a coordination number. (d) A proton flip within a chain reaction along two water molecules leads to the formation of one hydroxyl with the surface oxygen atom and a second surface hydroxyl upon dissociation of an initial molecular adsorbed water molecule. }
\label{pulling}
\end{figure}

The formation of a dense water layer at the surface promotes quite substantial (albeit local) rearrangement of the surface features, the most evident effect being the protruding of O atoms from their original positions towards the liquid solvent.
The height of these O atoms with respect to the surface plane changes by as much as 1.5\,\AA, as shown in three representative examples in Figure~\ref{pulling}.
As a result, the O atoms break one of their original bonds with Ti atoms underneath, decreasing their coordination number from 3 to 2, and instead become involved in direct hydrogen bonds with either chemisorbed or physisorbed water molecules of the first hydration cell.
Moreover, the transfer of a proton from one to another acceptor site on the surface was frequently observed, in some cases proceeding via a Grotthuss mechanism involving the adsorbed water molecules.
Effectively, the proton transfers led to a redistribution of the terminal OH and OH$_2$ groups over the surface, consolidating the hydrogen-bond network in immediate surface proximity.
Over the course of our short simulation, reactions of this kind took place especially at the bottom of the valleys, as already observed for the (100)-rec surface, due to the high density of surface terminal groups (O, OH, OH$_2$) and physisorbed H$_2$O molecules in those regions.

\begin{figure}[H]
\centering
\includegraphics[scale = 1]{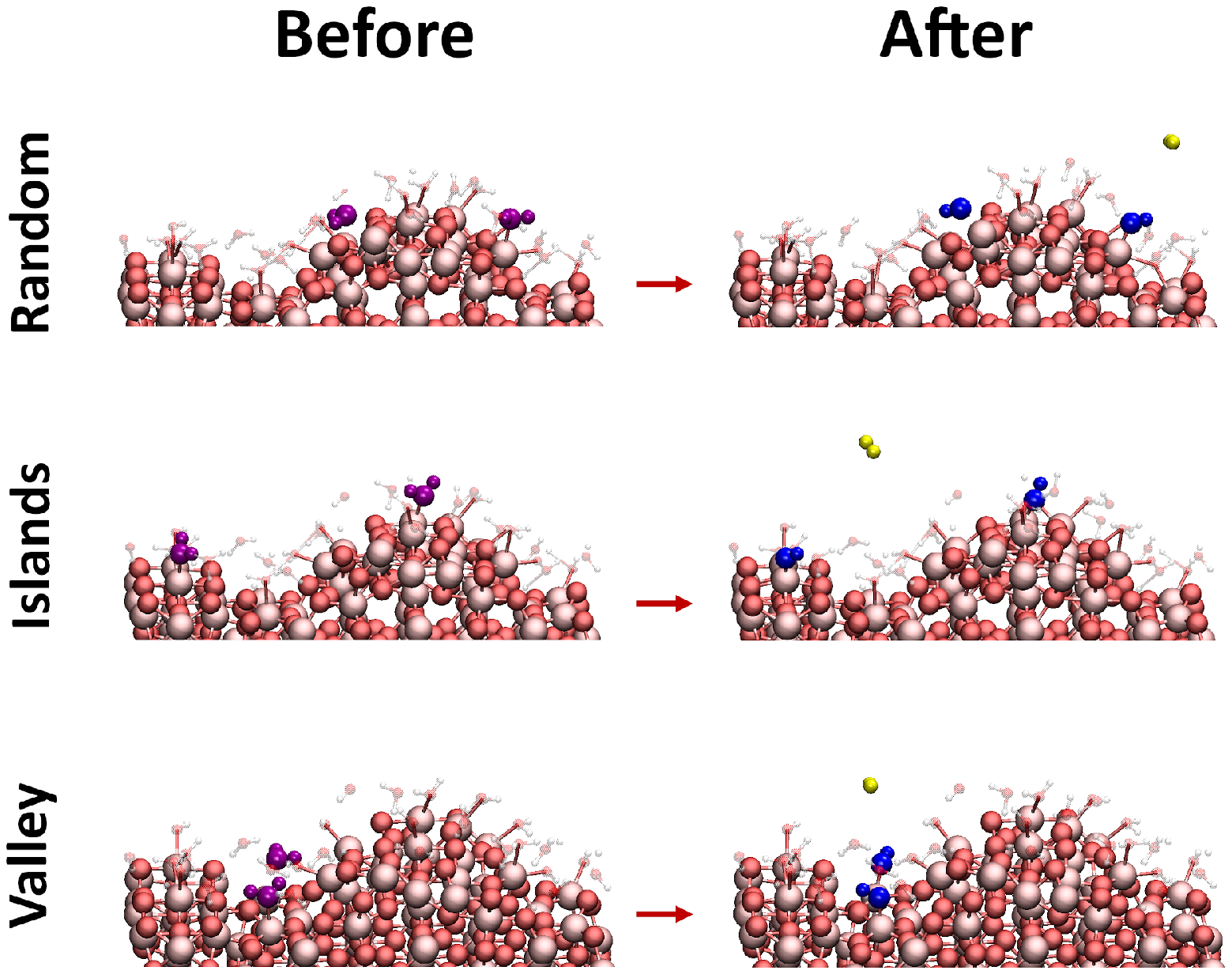}
\caption{Three different scenarios are investigated for the process of water splitting and the formation of molecular H$_2$ upon reaction of adsorbed  H$_2$O molecules on the reactive reconstructed (001) surface. The surface is shown as ball and stick model with metallic red oxygen and pink titanium atoms. The first adsorbed water layer is illustrated as small transparent ball and stick model, whereas the according water molecules chosen for the formation of molecular hydrogen are highlighted with the color code analogous to Figure \ref{all_h2o}. From the top to the bottom water molecules are chosen in random, islands or valley positions: (before) the chosen molecular waters before the suggested reaction turn to (after) surface hydroxyls and molecular hydrogen in its geometry optimzed configuration.}
\label{all_Geo_Stairs_AfterMD}
\end{figure}

Whereas only reactions involving transfers of protons between terminal sites were observed to occur spontaneously on the (001)-rec surface during our short MD simulation, the presence of excess Ti interstitials may promote thermally activated redox reactions with development of H$_2$, as a consequence of direct surface oxidation by adsorbing water molecules~\cite{Migani, Beinik}, according to
\begin{equation}
    2\text{H}_2\text{O}_{\text{(aq)}} + \text{Ti}_7\text{O}_{12} \longrightarrow 2\text{OH}^{-}_{\text{(ads)}} + \text{H}_{2\text{(aq)}} + \text{Ti}_7\text{O}^{\; \; 2+}_{12}
    \label{reaction}
\end{equation}
To test whether such a reaction may in principle take place, we compared the total energy of the (001)-rec model with only its first hydration shell (as obtained after geometry optimization at the end of the MD run described above) and of three other systems in which two protons were removed from adsorbed H$_2$O molecules while a H$_2$ molecule was placed in the vacuum space between the periodically repeated surface slabs (Figure~\ref{all_Geo_Stairs_AfterMD}).
In the three systems the two protons were chosen pseudo-randomly: from water adsorbed in the valleys, on top of the islands and from the islands' slopes.
Only in one case (proton removal from valley sites) the total energy after formation of the H$_2$ molecule increased by about 0.1~eV.
In the other two cases, the total energy {\em decreased} by -2.8\,eV (slope sites) and -3.0\,eV (islands' top sites).
After correcting those values for the Gibb's free energy of solvation of H$_2$ from a reference gas state at 1~atm to bulk liquid water at 300\,K, which amounts to +6.6\,kcal/mol~\cite{Smiechowski,Wilhelm_1977} (about +0.3\,eV), a driving force of at least -2.5\,eV can be estimated for the H$_2$ development reaction.
This suggests that water splitting with development of hydrogen gas can be promoted by TiO$_2$ surfaces in the presence of partially reduced Ti ions.
Whether similar mechanisms play a role during photocatalytic water splitting after UV irradiation of titanium oxide remains of course to be investigated.

\section{Conclusions}

Our comparative analysis of structural and energetic features of dry and wet rutile TiO$_2$ surfaces by means of DFT and DFTB shows that the latter formalism can be trusted to predict structural features (cell parameters, bond length and angles) of stoichiometric systems and static TiO$_2$/water interfaces.
However, it tends to overestimate the values of surface energies and the electron affinity of surface O species, even with the Tiorg-smooth parameter set.  
This, combined with the general tendency of overstabilizing hydrogen bonds, leads to prediction of predominantly dissociated surface hydration layers, especially in unbiased dynamical simulations, as already observed for other oxide systems such as ZnO~\cite{Svea}.
This is in contrast with GGA-PBE DFT, which predicts molecular adsorption to be favoured both at very low coverages and in the presence of bulk liquid water on most of the studied interfaces.
Exceptions occur in the presence of peculiar in-surface hydrogen-bonded patterns such as on the (110) facet at the water coverage of 1 monolayer, and whenever Ti atoms are strongly undercoordinated (fourfold or less), such as on the pristine (001) facet.
A certain amount of dissociative adsorption does also occur on the most complex and so far least studied reconstruction, namely the non-stoichiometric (001)-rec model.

The investigation of chemical reactions on surfaces of such conformational and chemical complexity, which can be modelled under periodic boundary conditions only with very large unit cells and include Ti atoms with oxidation states lower than 4, would benefit very strongly from the availability of accurate semiempirical formalisms such as DFTB.
Unfortunately, however, the current parameter sets fail in predicting the correct range of thermodynamic stability of this surface and also to reproduce the structural atomic arrangement obtained by DFT.
In order to overcome the problems experienced with the Tiorg and Tiorg-smooth SK-sets, a new DFTB parametrization is being currently developed~\cite{Verena_nanocrystals}. It offers a better description of the electronic part of the Ti-O interaction as it uses the more recent 3ob organic set \cite{3ob} for the description of the oxygen atoms. Additionally, it introduces three-body terms in the repulsive energy, resulting in an improved representation of undercoordinated Ti atoms on various TiO$_2$ surfaces.  
Until this new parametrization is completed and validated, in the present study we are limited to the rather short simulation time (5 to 10\,ps) accessible to DFT to understand the chemical behaviour of this surface in an aqueous environment.

Standard DFT does present the well-known inconvenience of strongly underestimating the band gap of all surface models, but is able to predict the predominantly molecular adsorption at very low coverage in line with all experimental findings~\cite{Diebold, Pang, Bourikas, Wu2}.
It also predicts the transition temperature towards the non-stoichiometric (001) reconstruction in remarkable agreement with the available experimental literature.~\cite{Tero,Noerenberg}.
We thus trust the formalism regarding its ability to predict the correct chemical behaviour of this system, which is interesting because of the presence of excess Ti interstitials and thus an overall reduced character with respect to bulk TiO$_2$.

Spontaneous reactions in a liquid environment are limited to dissociative adsorption and proton-exchange among different surface sites.
However, with the help of static calculations with geometry optimization we suggest that this surface can promote the thermally-activated splitting of water and release of H$_2$ molecules.
This finding opens up the possibility that similar reactions may occur during photocatalytic water splitting on TiO$_2$ materials upon exposure to UV light and formation of under-oxidised Ti defects at the surface or in sub-surface sites.

\begin{acknowledgement}

This work has been supported by the Deutsche Forschungsgemeinschaft through the Research Training Group 2247, Quantum Mechanical Material Modeling - QM$^3$. Computational time has been provided by the North-German Supercomputing Alliance (HLRN). The authors are thankful to Eric Macke for fruitful discussions.

\end{acknowledgement}

\providecommand{\latin}[1]{#1}
\makeatletter
\providecommand{\doi}
  {\begingroup\let\do\@makeother\dospecials
  \catcode`\{=1 \catcode`\}=2 \doi@aux}
\providecommand{\doi@aux}[1]{\endgroup\texttt{#1}}
\makeatother
\providecommand*\mcitethebibliography{\thebibliography}
\csname @ifundefined\endcsname{endmcitethebibliography}
  {\let\endmcitethebibliography\endthebibliography}{}

\end{document}